\documentclass[a4paper,USenglish]{lipics-v2021}

\usepackage{mathabx}
\usepackage[ruled,linesnumbered,vlined,boxed,commentsnumbered]{algorithm2e}
\usepackage{hyperref}
\usepackage{comment}
\usepackage{color}
\usepackage{graphicx}
\usepackage{microtype}
\usepackage[section]{placeins}
\def\lf{\tiny}
\newcounter{mylinenumber}

\def\nnll{\refstepcounter{mylinenumber}\lf\themylinenumber}

\newcommand{\commentline}[1]{\hspace{1cm}\{ \textit{#1} \}}

\usepackage{url}
\usepackage{cite}
\definecolor{heraldBlue}{rgb}{0.0,0.0,0.8}
\definecolor{heraldRed}{rgb}{0.8,0.0,0.0}
\definecolor{heraldGray}{rgb}{0.4,0.4,0.4}
\definecolor{heraldBlack}{rgb}{0.0,0.0,0.0} 
\definecolor{heraldGreen}{rgb}{0.0,0.4,0.0} 

\newcommand{\citep}{\cite}

\def\NOTES{1}
\def\SAVESPACE{0}

\if \SAVESPACE 1

\fi

\if \NOTES 1
\newcommand{\lfnote}[1]{\textcolor{heraldBlue}{\small \bf [LF: #1]}}
\newcommand{\trnote}[1]{\textcolor{heraldBlue}{\small \bf [TR: #1]}}
\newcommand{\stnote}[1]{\textcolor{heraldBlue}{\small \bf [ST: #1]}}
\newcommand{\pknote}[1]{\textcolor{heraldBlue}{\small \bf [PK: #1]}}
\else
\newcommand{\lfnote}[1]{}
\newcommand{\trnote}[1]{}
\newcommand{\stnote}[1]{}
\newcommand{\pknote}[1]{}
\fi

\newcommand{\true}{\textit{true}}
\newcommand{\false}{\textit{false}}

\newcommand{\ignore}[1]{}

\newcommand{\cP}{\mathcal P}

\newcommand{\Lat}{\mathcal L}
\newcommand{\Nat}{\mathbb{N}}

\title{Accountability and Reconfiguration: Self-Healing Lattice Agreement}

\author{Luciano Freitas de Souza}{LTCI, T\'el\'ecom Paris, Institut Polytechnique de Paris}{lfreitas@telecom-paris.fr}{}{}
\author{Petr Kuznetsov}{LTCI, T\'el\'ecom Paris, Institut Polytechnique de Paris}{petr.kuznetsov@telecom-paris.fr}{}{}
\author{Thibault Rieutord}
{CEA LIST, Université de Paris-Saclay}{thibault.rieutord@cea.fr}{}{}
\author{Sara Tucci-Piergiovanni}
{CEA LIST, Université de Paris-Saclay}{sara.tucci@cea.fr}{}{}

\Copyright{L. Freitas de Souza, P. Kuznetsov, T. Rieutord, S. Tucci-Piergiovanni}

\authorrunning{L. Freitas de Souza, P. Kuznetsov, T. Rieutord, S. Tucci-Piergiovanni}

\nolinenumbers

\begin{document}



\funding{Luciano Freitas de Souza was supported by Nomadic Labs, and Petr Kuznetsov---by TrustShare Innovation Chair.}

\maketitle







\begin{abstract}
An \emph{accountable} distributed system provides means to detect deviations of system components from their expected behavior. 
It is natural to complement fault detection with a reconfiguration mechanism, so that the system could heal itself, by replacing malfunctioning parts with new ones.
In this paper, we describe a framework that can be used to implement a large class of accountable and reconfigurable replicated services. 
We build atop the fundamental lattice agreement abstraction lying at the core of storage systems and cryptocurrencies.    

Our asynchronous implementation of accountable lattice agreement ensures that every violation of consistency is followed by an undeniable evidence of misbehavior of a faulty replica.
The system can then be seamlessly reconfigured by evicting faulty replicas, adding  new ones and merging inconsistent states. 
We believe that this paper opens a direction towards asynchronous ``self-healing'' systems that combine accountability and reconfiguration.
\end{abstract}




  

\section{Introduction} 
\label{sec:intro}
There are two major ways to deal with failures in distributed computing:
\begin{description}

\item[Fault-tolerance:] we anticipate failures by investing into replication and synchronization, so that the system's correctness is not affected by faulty components. 
\item[Accountability:] we detect failures \emph{a posteriori} and raise undeniable evidences against faulty components. 

\end{description}
Accountability in computing has been proposed for generic distributed systems~\citep{detection-case,detection-problem} as a mechanism to detect deviations of system nodes from the algorithms they are assigned with.      
It has been shown that a large class of  deviations of a given process from a given deterministic algorithm can be detected by maintaining a set of \emph{witnesses} that keep track of all \emph{observable} actions of the process and check them against the algorithm~\citep{peerreview}.     

The generic approach can be, however, very expensive in practice and one may look for a more tractable, \emph{application-specific} accountability mechanism.
Indeed, instead of pursuing the ambitious goal of detecting deviations from the assigned algorithm, we might want to only care about deviations that violate the specification of the problem the algorithm is trying to solve. 

The idea has been successfully employed in the context of Byzantine Consensus~\citep{civit2021polygraph}.
The accountable version of consensus guarantees correctness as long as the number of faulty processes does not exceed some fixed $f$. 
But if correctness is violated, e.g., honest processes take different decisions, then at least $f+1$ Byzantine processes are presented with undeniable evidences of misbehavior.
This is not surprising: a decision in a typical $f$-resilient consensus protocol must receive \emph{acknowledgements} from a \emph{quorum} of processes, and any two quorums must have at least $f+1$ processes in common~\citep{PSL80}.
The fact that two processes took different decisions implies that at least $f+1$ processes in the intersection of the corresponding quorums \emph{equivocated}, i.e., acknowledged conflicting decision values. 
Assuming that every decision is provided with a cryptographic \emph{certificate} containing the set of signed acknowledgements from a quorum of processes, we can immediately construct a desired evidence.  
Polygraph~\citep{civit2021polygraph}, a recent accountable Byzantine Consensus protocol,  naturally builds upon the classical PBFT protocol~\citep{CL02}.   
One may ask---okay, we have detected a faulty process, but what should we do next?
Ideally, we would like to \emph{reconfigure} the system by evicting the faulty process and \emph{reinitializing} the system state. 

\emph{Reconfigurable replicated systems}~\citep{rambo,parsimonious,smartmerge,SKM17-reconf} allow the users to dynamically update the set of replicas.  
It has been recently shown that reconfiguration can be implemented in purely \emph{asynchronous}
environments~\citep{dynastore,parsimonious,smartmerge,freestore,SKM17-reconf,rla}.
The idea was first applied to (read-write) storage systems~\citep{dynastore,
  parsimonious,freestore}, and then extended to
max-registers~\citep{smartmerge,SKM17-reconf} and more general \emph{lattice} data types, first in the crash-fault context~\citep{rla} and then for Byzantine failures~\citep{rbla}.

\subparagraph*{Contribution.} In this paper, we propose a framework that can be used to implement a large class of replicated services that are both accountable and reconfigurable. 
Following recent work on reconfiguration~\citep{smartmerge,rla,rbla}, we build atop the fundamental \emph{lattice agreement} abstraction.
Lattice agreement~\citep{lattice-hagit,gla} (LA) takes arbitrary inputs in a \emph{lattice} (a partially ordered set equipped with a \emph{join} operator) and returns outputs that are (1)~joins of the inputs, and (2)~ordered with respect to the lattice partial order. 
%
The LA abstraction is weaker than consensus and can be implemented in an asynchronous system.   

Lattice agreement appears to be a perfect match for both desired features: accountability and reconfiguration.
Indeed, a quorum-based LA implementation enables detection of misbehaving parties: as soon as two correct users learn two incomparable values, they also obtain  a proof of misbehavior of all replicas that \emph{signed} both values. 
Furthermore, the very process of reconfiguration can be represented as agreement defined on a lattice of \emph{configurations}~\citep{smartmerge,rla}.
These two observations inspire the design of our system.

We propose an accountable \emph{and} reconfigurable implementation that reaches agreement on a \emph{joint} lattice: an object lattice (defining the current \emph{state} of the replicated object) and a configuration lattice (defining the current \emph{configuration} of the replicas).     
Assuming that the number of failures is less than half of the system size, our implementation is \emph{alive}.
It is also \emph{safe} if only benign (crash) failures occur.
Once safety is violated, i.e., two correct users learn two incomparable object states, some Byzantine replicas are inevitably confronted with an undeniable proof of misbehavior.
The system is then seamlessly reconfigured by evicting the detected replicas, adding  new ones and merging inconsistent states.
%
Once the state is merged, the system comes back to providing safety and liveness, as long as no new replicas exhibits Byzantine behavior. 
Eventually all Byzantine replicas are detected and the system comes back to maintaining both liveness and safety.

\subparagraph*{Outdated configurations are harmless.} Our system prevents users from accessing outdated configurations with the use of \emph{forward-secure digital signature scheme}~\citep{bellare1999forward,drijvers2019pixel}. 
%
A member of each new configuration is assigned a new secret key. 
Furthermore, honest members of an old configuration are expected to destroy their old keys before moving to a new one.
Thus, if they are later compromised, they will not be able to serve clients' requests, and the remaining Byzantine replicas will not constitute a quorum.  

\subparagraph*{On Byzantine clients.} For simplicity, our solution assumes that service replicas are subject to Byzantine failures, but clients are \emph{benign}: they can only fail by crashing.
This assumption has already been made in designs of fault-tolerant storage systems \citep{MAD02-storage}.
In our case, it precludes the cases when a Byzantine client brings the system into a compromised configuration or slows down the system by issuing excessive reconfiguration requests.  
In Appendix~\ref{sec:A1LA} we also describe a \emph{one-shot} version of accountable lattice agreement, without reconfiguration, in which both clients and replicas can be Byzantine.
Marrying reconfiguration and accountability in a \emph{long-lived} service that can be accessed by Byzantine clients remains an important challenge.
One way to address it is to assume an external \emph{access control} mechanism~\cite{bla} ensuring that only "authentic" configurations are accepted as inputs to the reconfiguration procedure.  
We discuss this issue in more detail in Section~\ref{sec:discussion}. 

\subparagraph*{Summary.} 
Altogether, we believe that this paper opens a new area of asynchronous ``self-healing''  systems that combine accountability and reconfiguration.
Such a system either preserves safety and liveness or preserves liveness and compensates safety violations with eventual detection of Byzantine replicas. 
It also exports a reconfiguration interface that allows the clients to replace compromised replicas with new, correct ones. 
In this paper, we show that both mechanisms, accountability and reconfiguration, can be implemented in a purely asynchronous (in the modern parlance---\emph{responsive}) way.

\subparagraph*{Road map.} The rest of the paper is organized as follows. 
In Section~\ref{sec:model}, we introduce our system model.
In Section~\ref{sec:rala}, we state the problem of reconfigurable and accountable lattice agreement (RALA) and in Section~\ref{sec:algo}, we describe our RALA implementation analysing its correctness.
In Section~\ref{sec:related}, we discuss related work, and in Section~\ref{sec:discussion} we present an overview of possible improvements and interesting open questions.  
In Appendix~\ref{sec:A1LA}, we present our  one-shot accountable lattice agreement (A1LA) that assumes that both clients and replicas can be Byzantine and analyse its correctness.    
\FloatBarrier

\section{System Model}
\label{sec:model}
%
We assume that the system is asynchronous and that it is composed by a set $\Pi$ of processes that communicate over reliable message-passing channels
exchanging authenticated messages. These processes are split into a set $\Sigma$ of \emph{replicas} that maintain a \emph{replicated service} and a set $\Gamma$ of \emph{clients} that use the service.
We assume the existence of a global clock with range $\Nat$, but the processes do not have access to it. 

In each run, a process can be: 
(1)~\emph{correct} ($C$) if it faithfully follows the algorithm it is assigned with, (2)~\emph{benign} ($B$) if it can only deviate from the algorithm by prematurely stopping taking steps of its
algorithm, or  (3)~\emph{malicious} ($M$) or Byzantine if it skips steps or takes
a step not prescribed by its algorithm.   


We assume a \emph{forward-secure digital signature scheme}~\citep{bellare1999forward,malkin2002efficient,boyen2006forward,drijvers2019pixel}.
%
%
In the scheme, the public key of a process $p$ is fixed while its secret key $sk^p_{t}$ evolves with its \emph{timestamp} $t$, a natural number bounded by a fixed natural parameter $T$, usually taken sufficiently large  (e.g., $2^{64}$), to accommodate any possible system lifetime. 
For any $t'$, $t< t' \le T$, the process can \emph{update its secret key} and obtain $sk^p_{t'}$ from $sk^p_{t}$.
However, we assume that it is computationally infeasible to ``downgrade'' the key to a lower timestamp, from $sk^p_{t'}$ to $sk^p_{t}$.
In particular, once a process updates its timestamp from $t$ to $t'>t$, and then destroys $sk^p_t$, it is no longer able to sign messages with timestamp less than $t'$, even if it turns Byzantine later.    

More formally, we model a forward-secure signature scheme as an oracle which associates every process $p$ with a timestamp $t_p$. 
The oracle provides process $p$ with three operations:
(1) $\mathit{UpdateFSKeys(t)}$ sets $t_p$ to $t$ if $t$ is greater than the current value of $t_p$ but less or equal to $T$
(2) $\mathit{FSSign(m,t)}$ returns a signature $s$ for message $m$ and timestamp $t$, assuming $t \geq t_p$;
(3) $\mathit{FSVerify(m,t,s,q)}$ returns \textit{true} iff the message $m$ provides a signature $s$ generated by a valid call $\mathit{FSSign(m,t)}$ by
        process $q$.

We also make use of a weak broadcast primitive that ensures that once a correct process broadcasts a message, all correct processes eventually receive it, e.g., via a gossip mechanism.
Notice that, unlike reliable broadcast~\cite{Bra87a,rsdp-book2011}, we only require the primitive to disseminate messages broadcast by correct processes, not to make them eventually agree on the set of delivered ones. 

We assume that all clients are benign. 
For the sake of simplicity, we assume that once a correct process learns an output, it eventually  proposes a new input,
and that there are only finitely many correct clients.\footnote{Our specification can be easily refined to accommodate infinitely many correct clients under the assumption that the number of \emph{concurrently} proposed values is bounded.} 
%
\FloatBarrier

\section{Reconfigurable and Accountable Lattice Agreement: Specification}
\label{sec:rala}

A lattice is a partially ordered set where any pair of elements has a unique \emph{join}, or supremum, and a unique \emph{meet}, or infinum.  
We denote $\mathcal{O}$ the object lattice corresponding to the data type the user wishes to implement using the system (such as a counter, a set or commit-abort) and $K$ the configuration lattice. 

A \emph{configuration} $\kappa$ is a finite set of pairs 
$(\sigma, \mathit{inout}) \vert \sigma \in \Sigma, \mathit{inout} \in \{+,-\}$.  
Intuitively, $(\sigma,+)\in\kappa$ means that $\sigma$ has been earlier added to the configuration and $(\sigma,-)$ means that $\sigma$ has been removed from it.
We say that a replica $\sigma$ is a member of $\kappa$ if 
$(\sigma,+) \in \kappa$ and 
$(\sigma,-) \notin \kappa$.


$\lvert \kappa \rvert$ is defined as the cardinality of the set of pairs representing the configuration; $\kappa.\mathit{excluded}$ returns all the replicas excluded from it; $\kappa.\mathit{included}$ that were at some moment included on it; $\kappa.\mathit{members} := \kappa.\mathit{included}\backslash \kappa.\mathit{excluded}$.
We only consider \emph{well-formed} configurations $\kappa$: $\kappa.\mathit{excluded}\subseteq \kappa.\mathit{included}$ (a replica can be removed only if it has been previously added). 

In the \emph{reconfigurable accountable (long-lived) lattice agreement} (\emph{RALA}) abstraction, defined on a product lattice
$(\Lat, \sqsubseteq) = (\mathcal{O} \times K, \sqsubseteq^\mathcal{O} \times \sqsubseteq^K$), a client $c_i$ periodically proposes inputs $(\iota, \kappa) \vert \iota \in \mathcal{O}, \kappa \in K$ to replicas in
$\Sigma$ and obtains, as output, a value $\upsilon\in \Lat$.

Additionally, the client locally maintains an \emph{accusation set}  $\alpha_i=(A,P)$ where $A \subset \Sigma$ is a set of replicas and $P\in \cP$ is a \emph{proof}
(here $\cP$ is the set of proofs).
The system provides a Boolean map $\textit{verify-proof}: (2^{\Pi}\times \cP)\to \{\true,\false\}$ that can be used by any process or third party to \emph{verify} a proof. 
For example, a proof can be a set of messages that, for every replica in $r\in A$, contains one or more messages signed by $r$ that cannot be sent by $r$ in any execution of our algorithm.

When a client $c$ receives an input $v$ from the upper-level application we
say that $c$ \emph{proposes} $v$.
When  $c$ outputs a value $v\in\Lat$,
we say that $c$ \emph{learns} (or \emph{decides}) $v$.
When $c$ sets its accusation set to $(A,P)$, we say that $c$ \emph{accuses $A$ with~$P$}.

Given a client $c_i$, let $I^{i} = \langle (\iota^{i}_0, \kappa^i_0), (\iota^{i}_1, \kappa^i_1), \cdots \rangle$
denote the sequence of inputs and $\Upsilon^{i} = \langle \upsilon^{i}_0, \upsilon^{i}_1, \cdots \rangle$
denote the sequence of outputs.
%
If for some client $c_i$ and $k\in\Nat$, $\kappa^i_k\neq\kappa^i_{k+1}$, i.e., $c_i$ proposes to change the configuration, we say that $c_i$  \emph{issues a reconfiguration request}. 

Now a \emph{RALA} system must ensure the following properties:

\begin{itemize}
\item {\bf Validity.} Each value $\upsilon^{i}_k$, $k\geq 0$, learned by a client $c_i$
   is a join of the $k$-prefix of its input sequence and some values from other client's inputs. 

\item {\bf Completeness.} If a correct client learns a value that is incomparable with a value learnt by another correct client then it eventually accuses some replicas it had not yet accused before. 
    \[
        \forall c_i, c_j \in C\cap \Gamma, \forall k,l \in \mathbb{N}, \lnot \left( \upsilon^i_{k} \sqsubseteq
        \upsilon^j_{l} \lor \upsilon^j_{l} \sqsubseteq \upsilon^i_{k} \right) \mbox{, where $c_i$ learns  $\upsilon^i_{k}$ at time $t$} 
        \]
        \[
        \implies \exists t'>t:\; A^i[t] \subsetneq A^i[t']
    \]

\item {\bf Accusation Stability.} The accusation sets monotonically increase.
    \[
        \forall c_i \in \Gamma, t,\;t'\in\mathbb{N}, t<t': A^i[t] \subseteq A^i[t']
    \]

\item {\bf Accuracy.} If a client accuses a set of replicas $A$, then it has a valid proof against each replica in $A$:
    \[
        \forall c_i \in \Gamma, \forall t \in \mathbb{N}, \textit{verify-proof}(A^i[t],P^i[t])
    \]
    
\item {\bf Authenticity.} It is computationally infeasible to accuse a benign process, i.e., to construct $P\in\cP$ s.t.  $\textit{verify-proof}(A,P)$ $=\true$ and $A\cap B \neq \emptyset$.      

\item {\bf Agreement.} The correct clients eventually agree on the replicas they accuse.
    \[
        \forall t \in \mathbb{N}, \; \forall c_i, c_j \in \Gamma,\; \exists t' \in \mathbb{N}, \; t'>t:\;  A^i[t] \subseteq A^j[t']
    \]

\item {\bf Liveness.} If the system reconfigures only finitely many times, every value
    proposed by a correct client is eventually included in the value learned by every correct client.
    \[
        \forall c_i \in \Gamma, \forall k \in \mathbb{N}, \forall c_j \in \Gamma, \exists \ell \in \mathbb{N} \vert \iota^i_k \sqsubseteq \upsilon^j_l
    \]
  
\end {itemize}

A configuration $\kappa$ is said to be \emph{active} (at a given moment of time $t$) if (1)~it is a join of configurations proposed and learnt by time $t$, (2)~and no other correct process learns a configuration $\kappa ' \vert \kappa \sqsubset \kappa '$ by time $t$.
Liveness guarantees of our algorithm rely upon the following condition:
\begin{description}
\item[Configuration availability:] 
For all times $t$, any configuration that is 
active at all $t'>t$ contains a majority of correct processes.
\end{description}
This is a conventional assumption in asynchronous reconfigurable systems~\cite{dynastore,SKM17-reconf,rla}.
The intuition behind it is the following. If an active configuration remains active forever, i.e., it is never superseded, then it should contain enough correct replicas. 
On the other hand, a once active but later superseded configuration may contain arbitrarily many Byzantine processes: the clients' requests will be served by the new configuration.      


Notice that the properties above imply that either the values learnt by correct processes are comparable or eventually some Byzantine replicas are detected.   
If from some point on, no more Byzantine faults take place, we ensure that all new learnt values are comparable.  
Our requirement of finite number of reconfigurations is  standard in the corresponding literature~\cite{SKM17-reconf,rla,freestore}
and, in fact, can be shown to be necessary~\cite{r-liveness}.
In practice, we ensure liveness in ``sufficiently long'' time intervals without reconfiguration.   

Notice that the choice of new configurations to propose is left entirely to the clients, as long as the condition above is satisfied.  
In Section~\ref{sec:discussion}, we discuss possible reconfiguration strategies the clients may want to choose. 
However, it is important to emphasize that regardless of this strategy, \textbf{the system does not allow the accused replicas to affect the system's safety and liveness anymore}.


%
\FloatBarrier

\section{Reconfigurable and Accountable Lattice Agreement: Implementation}
\subsection{Algorithm}
\label{sec:algo}
%
Our RALA implementation is given in \autoref{fig:alapart1}, \autoref{fig:alapart2}, and \autoref{fig:alapart3}. 
We assume that every method in the  algorithms is executed by the process \emph{sequentially}, without being interrupted by other methods of this process. 
Moreover, we consider that the processes \emph{ignore} accused replicas, messages with invalid signatures and messages whose signatures do not match the configuration content.

\begin{figure}[!hb]
\hrule
 {\small
\begin{tabbing}
 bbbb\=bbbb\=bbbb\=bbbb\=bbbb\=bbbb\=bbbb\=bbbb \=  \kill
\textbf{operation} \textit{Verify-Proof(accusation(A,P))} \\
\nnll\label{line:vp:loopProcess}\> \textbf{foreach} Process $b \in A$ \textbf{do}\\
\nnll\label{line:vp:groupMessages}\>\> \textbf{let} $MSG$ be the union of all messages by $b$
in $\mathit{P}$\\
\nnll\label{line:vp:checkSigns}\>\>\> Check if every $m$ in $MSG$ has a valid signature \textbf{continue} if not\\
\nnll\label{line:vp:checkACKs}\>\>\> Get all ACKs in $MSG$ and check if they are comparable,
\textbf{continue} if not\\
\nnll\label{line:vp:checkInputs}\>\>\> Get all Proposal in $MSG$ and check if they obey the description \textbf{continue} if not\\
\nnll\label{line:vp:checkDecision}\>\>\> Get all Decision in $MSG$ and check if their ACKs hold
\textbf{continue} if not\\
\nnll\label{line:vp:innocent}\>\>\> \textbf{return} \textit{false}\\
\nnll\label{line:vp:guilty}\> \textbf{return} \textit{true}
\end{tabbing}
}
\vspace{-1.5mm}
 \hrule
\caption{Example of Verify-proof Operation}
\label{fig:verify-proof}
\end{figure}

\begin{figure}[!hb]
\hrule \vspace{1mm}
 {\small
 \setcounter{linenumber}{0}
\begin{tabbing}
 bbbb\=bbbb\=bbbb\=bbbb\=bbbb\=bbbb\=bbbb\=bbbb \=  \kill
\textbf{Local variables:} \\
\textit{status}, initially waiting~~\{ Boolean indicating status: waiting or proposing \}\\
\textit{dest}, initially $\emptyset$~~\{ Set of replicas that must be contacted \}\\
\textit{nackBool}, initially false~~\{ Flag indicating whether a NACK has already been received or not \}\\
\textit{activePropNb}, initially $-1$~~\{ Index of the current active proposal \}\\
\textit{activeOutNb}, initially $0$~~\{ Index of the next value to be learnt \}\\
\textit{propV}, initially $\bot$~~\{ Value currently being proposed \}\\
\textit{objL}, initially empty~~\{ Ledger matching object values in the system to their original proposer \}\\
\textit{confL}, initially containing \textit{Initial Conf} signed by $c$~~\{ Analogous to objL for configurations \}\\
\textit{ackL}, initially empty~~\{ Ledger matching acks to the replicas that issued them \}\\
\textit{pendConf}, initially $\emptyset$~~\{ Set of pending configurations \}\\
\textit{RESPSet}, initially $\emptyset$~~\{ Set of replicas that responded \}\\
\textit{lastDec}, initially $(\bot, intialConfig)$~~\{ Last decided value \}\\
\textbf{Input:}\\
\textit{inBuffer} ~~ \{ Values received by the client from an external source to insert in the system \}\\
\textbf{Outputs:}\\
\textit{outV}, initially $\bot$~~\{ Array of values learnt by the client \}\\
\textit{accusation}, initially $\emptyset$~~\{ Set of accusations issued by the client \} \\
\\
\textbf{upon} $\mathit{status} = \mathit{waiting}$ AND $\mathit{inBuffer} \neq \bot$\\
\nnll\label{line:ala:setobjL}\> \textit{extract} and \textit{sign} objects from $\mathit{inBuffer}$ and \textit{include} them to $\mathit{objL}$\\
\nnll\label{line:ala:setconfL}\> \textit{extract} and \textit{sign} configurations from $\mathit{inBuffer}$ and \textit{include} them to $\mathit{confL}$\\
\nnll\label{line:ala:callPropose}\> Propose \\
\\
\textbf{operation} \textit{Propose} \\
\nnll\label{line:ala:getProposedValue}\> $\mathit{propV} := \mathit{extractLedger(objL, confL, c)}$\\
\nnll\label{line:ala:pending}\> include $\mathit{propV.conf}$ to $\mathit{pendConfSet}$\\
\nnll\label{line:ala:activation}\> $\mathit{status} := \mathit{proposing}$\\
\nnll\label{line:ala:incrementProposal}\> $\mathit{activePropNb} := 
\mathit{activePropNb} + 1$\\
\nnll\label{line:ala:scrapAcks}\> clear \textit{ackL[activeOutNb]} \\
\nnll\label{line:ala:scrapRESP}\> clear \textit{RESPSet} \\
\nnll\label{line:ala:counterReset}\> $\mathit{nackBool} := false$\\
\nnll\label{line:ala:getDest}\> $\mathit{dest := propV.conf.included - lastDec.conf.excluded}$\\
\nnll\label{line:ala:mcastProposal}\> \textit{multicast} $\langle \textit{PROPOSAL}, 
(objL, confL, lastDec, activePropNb) \rangle$ to replicas in \textit{dest}\\
\\
\textbf{upon} $\mathit{verify\_maj}(RESPSet, lastDec.conf, pendConf) = \mathit{true}$ \\
\nnll\label{line:ala:checkNack}\> \textbf{if} $\mathit{nackBool} = true$ \textbf{then} Propose \textbf{else} Decide\\
\\
 \textbf{upon} \textit{receive} $\langle \textit{ACK}, (HASH(propV), lastDec, pendConf', \mathit{activePropNb}) \rangle$ \\
\>from replica $r$ AND $\mathit{status} = \mathit{proposing}$ AND $r \not \in ackL$ AND $r \in \mathit{dest}$\\
\nnll\label{line:ala:noPendingCheck}\> \textbf{if} $\mathit{propV.conf \in pedingConf'}$ \textbf{then}\\
\nnll\label{line:ala:appendACKList}\>\> append $r$'s ACK message to $\mathit{ackL[activeOutNb]}$\\
\nnll\label{line:ala:includeACKPending}\>\> include elements from $\mathit{pendConf'}$ which aren't subset of $\mathit{lastDec.conf}$ in $\mathit{pendConfSet}$\\
\nnll\label{line:ala:appendACKRESPList}\>\> append $r$ to $\mathit{RESPSet}$\\
\nnll\label{line:ala:elseAccuseACK}\> \textbf{else} $\langle \textit{ACCUSATION},
(accusation) \rangle$ \\
\nnll\label{line:ala:noPendingConv}\>\> include $(r,ACK)$ to \textit{accusation}\\
\nnll\label{line:ala:broadcastByzantinePending}\>\> \textit{broadcast} $\langle \textit{ACCUSATION},
(accusation) \rangle$
\end{tabbing}
}
\vspace{-1.5mm}
 \hrule
\caption{Reconfigurable Accountable Lattice Agreement: Code for client $c$ part 1.}
\label{fig:alapart1}
\end{figure}

\begin{figure}[!hb]
\hrule \vspace{1mm}
 {\small
\begin{tabbing}
 bbbb\=bbbb\=bbbb\=bbbb\=bbbb\=bbbb\=bbbb\=bbbb \=  \kill
\textbf{upon} \textit{receive} $\langle \textit{NACK}, (HASH(propV), \mathit{\Delta objL}', \mathit{\Delta confL}', \mathit{activePropNb}) \rangle$ from replica $r$\\
\> AND $\mathit{status} = \mathit{proposing}$ AND $r \in \mathit{dest}$\\
\nnll\label{line:ala:getDeltaValue}\>  $\mathit{nackV} := \mathit{extractLedger}(\mathit{\Delta objL}',\mathit{\Delta confL}', r)$\\
\nnll\label{line:ala:nonDelta}\> \textbf{if} $\mathit{nackV} \sqsubseteq \mathit{propV}$  \textbf{return}\\
\nnll\label{line:ala:updateobjLProposal}\> $\mathit{objL := objL \cup objL'}$\\
\nnll\label{line:ala:updateconfLProposal}\> $\mathit{confL := confL \cup confL'}$\\
\nnll\label{line:ala:incrementNACKCount}\> $\mathit{nackBool} := true$ \\
\nnll\label{line:ala:appendNACKRESPList}\> append $r$ to $\mathit{RESPSet}$\\
\\
\textbf{operation} \textit{Decide} \\
\nnll\label{line:ala:updateOutput}\> $\mathit{outV[activeOutNb]} := \mathit{propV}$ \\
\nnll\label{line:ala:bcastDecision}\> \textit{broadcast} $\langle \mathit{DECISION}, 
(\mathit{objL, confL}, \mathit{ackL[activeOutNb]}) \rangle$\\
\nnll\label{line:ala:updateDecided}\> $\mathit{lastDec} := \mathit{outV[activeOutNb]}$\\
\nnll\label{line:ala:excludeSuperseeded}\> $\mathit{pendConfSet} := \emptyset$ \\
\nnll\label{line:ala:incrementOutput}\> $\mathit{activeOutNb} := 
\mathit{activeOutNb} + 1$\\
\nnll\label{line:ala:decidePassive}\> $\mathit{status} := \mathit{waiting}$ \\
\\
\textbf{upon} \textit{receive} $\langle \textit{DECISION}, (\mathit{objL}', \mathit{confL}', \mathit{ackL}') \rangle$ from client $c'$\\
\nnll\label{line:ala:getoutV}\> $\mathit{outV}' := \mathit{extractLeger}(\mathit{objL}', \mathit{confL}')$\\
\nnll\label{line:ala:lastBCKUP}\> $\mathit{lastDecOld} := \mathit{lastDec}$\\
\nnll\label{line:ala:updateConfRec}\> $\mathit{lastDec} := \mathit{lastDec} \cup \mathit{outV'}$\\
\nnll\label{line:ala:excludeSuperseededRec}\> Eliminate from $\mathit{pendConf}$ subsets of $\mathit{lastDec.conf}$\\
\nnll\label{line:ala:completeMissedobjL}\> $\mathit{objL} := \mathit{objL}' \cup \mathit{objL}$\\
\nnll\label{line:ala:completeMissedconfL}\> $\mathit{confL} := \mathit{confL}' \cup \mathit{confL}$\\
\nnll\label{line:ala:checkUncomparability}\> $\forall i ~\vert ~
\mathit{outV}' \not \sqsubseteq
\mathit{outV[i]} ~\&\&~ \mathit{outV[i]} \not \sqsubseteq \mathit{outV}'$ \\
\nnll\label{line:ala:pickByzantineAcker}\>\> \textbf{let} $M = \{m \vert m \in \mathit{ackL[i]} ~\&\&~ m\in \mathit{ackL}' ~\&\&~ m \not \in accusation\}$\\
\nnll\label{line:ala:proofAcker}\>\> \textbf{foreach} $m \in M$ \textbf{do}
~include $(m, \{\mathit{ackL}[m], \mathit{ackL}'[m])\})$ to \textit{accusation}\\
\nnll\label{line:ala:byzantineAckerPassive}\>\> \textbf{if} $\lvert M \rvert > 0$ \textbf{then} \textit{broadcast} $\langle \textit{ACCUSATION},
(accusation) \rangle$\\
\nnll\label{line:ala:checkReproposql}\> \textbf{if} $\mathit{lastDec.conf \not\sqsubseteq lastDecOld.conf \lor outV' \not\sqsubseteq propV}$ \textbf{then} Propose\\
\\
\textbf{operation} \textit{extractLedger} $(\mathit{objL}', \mathit{confL}', \mathit{sender})$\\
\nnll\label{line:ala:checkByzantineSender}\> \textbf{if} $\exists~\mathit{process}~p \in \mathit{objL}'$ or $\mathit{confL}'$ with invalid signature \textbf{then}\\
\nnll\label{line:ala:convictSender}\>\> $\mathit{accusation} := \mathit{accusation}
\cup \{(\mathit{sender}, \mathit{getMSG(objL') \cup \mathit{getMSG(confL')}}\}$\\
\nnll\label{line:ala:convictSSenderpassive}\>\> \textit{broadcast} $\langle \textit{ACCUSATION},
(accusation) \rangle$\\
\nnll\label{line:ala:convictSSenderreturn}\>\> \textbf{return} $\emptyset$\\
\nnll\label{line:ala:getInputValue}\> \textbf{let} $\mathit{receivedValue} = (\sqcup [v \vert \exists p, \mathit{objL}'[p] = v], \sqcup [c \vert \exists p, \mathit{confL}'[p] = c])$\\
\nnll\label{line:ala:ledgerReturn}\>\> \textbf{return} $\mathit{receivedValue}$\\
\\
\textbf{upon} \textit{receive} $\langle \textit{ACCUSATION}, \mathit{(accusation')} \rangle$ from client $q$\\
\nnll\label{line:ala:initDeltaProof}\> $\mathit{\Delta Proof} := \emptyset$ \\
\nnll\label{line:ala:loopConviction}\> \textbf{foreach} process $b$ accused in \textit{accusation'} 
with $p$ and who isn't present in \textit{accusation} \textbf{do}\\
\nnll\label{line:ala:newConvictionFound}\>\> include $(b,p)$ in $\mathit{\Delta Proof}$\\
\nnll\label{line:ala:checkDelta}\> \textbf{if} $\Delta Proof \neq \emptyset$ \textbf{then} \\
\nnll\label{line:ala:addProof}\>\> $\mathit{accusation} := \mathit{accusation}
\cup \Delta Proof$
\end{tabbing}
}
\vspace{-1.5mm}
 \hrule
\caption{Reconfigurable Accountable Lattice Agreement: Code for client $c$ part 2.}
\label{fig:alapart2}
\end{figure}

\begin{figure}[!hb]
\hrule \vspace{1mm}
 {\small
\begin{tabbing}
 bbbb\=bbbb\=bbbb\=bbbb\=bbbb\=bbbb\=bbbb\=bbbb \=  \kill
\textbf{Local variables:} \\
\textit{objL}, initially empty  \{ Object Ledger \}\\
\textit{confL}, initially empty \{ Configuration Ledger \}\\
\textit{repV} initially $\bot$ \{ Value held by replica \}\\
\textit{pendConf}, initially $\emptyset$ \{ Pending Configurations \}\\
\textit{lastDec}, initially $\bot$\\
signature timestamp $\mathit{t_r}$ initially |\textit{Initial Configuration}|\\
\\
\textbf{sign} all outgoing messages $m$ with \textit{FSSign}$(m, t_r)$ \\
\\
\textbf{upon} \textit{receive} $\langle \textit{PROPOSAL}, (\mathit{objL}', \mathit{confL}', \mathit{lastDec'}, \mathit{activePropNb}') \rangle$ from client $c$\\
\nnll\label{line:ala:checklastDec}\> \textbf{if} $\mathit{lastDec' \sqsubseteq lastDec}$ \textbf{then return}\\
\nnll\label{line:ala:updatelastDec}\> $\mathit{lastDec := lastDec \cup lastDec'}$\\
\nnll\label{line:ala:getProposal}\> $\mathit{propV'} := \mathit{extractLedger}(\mathit{objL}',\mathit{confL}')$\\
\nnll\label{line:ala:updateServobjL}\> $\mathit{objL := objL \cup objL'}$\\
\nnll\label{line:ala:updateServconfL}\> $\mathit{confL := confL \cup confL'}$\\
\nnll\label{line:ala:checkInclusion}\> \textbf{if} $\mathit{repV} \sqsubseteq \mathit{propV}'$ \textbf{then}\\
\nnll\label{line:ala:updateAccepted}\>\> $\mathit{repV} := \mathit{propV}'$\\
\nnll\label{line:ala:includePending}\>\> Include $\mathit{propV'.conf}$ to \textit{pendConf}\\
\nnll\label{line:ala:rcaulculatest}\>\> $t_r := $ |\textit{repV.conf}|\\
\nnll\label{line:ala:serverforward}\>\> $\mathit{UpdateFSKeysDestroyOld(t_r)}$\\
\nnll\label{line:ala:sendACK}\>\> send $\langle \textit{ACK}, (HASH(propV'), lastDec, pendConf, \mathit{activePropNb}') \rangle$ to $c$\\
\nnll\label{line:ala:newValueSeen}\> \textbf{else} \\
\nnll\label{line:ala:updateNAccepted}\>\> $\mathit{repV} := \mathit{repV} \sqcup \mathit{propV}'$\\
\nnll\label{line:ala:nackknewValueSeen}\>\> send $\langle \textit{NACK}, (HASH(propV'), \mathit{objL-objL'},\mathit{confL-confL'}, \mathit{activePropNb}') \rangle$ to $c$\\
\\
\textbf{upon} \textit{receive} $\langle \textit{DECISION}, (\mathit{objL}', \mathit{confL}', \mathit{ackL}') \rangle$ from client $c$\\
\nnll\label{line:ala:rupdateConfRec}\> $\mathit{lastDec} := \mathit{lastDec} \cup$ \textit{extractLedger} $(\mathit{objL}', \mathit{confL}')$\\
\nnll\label{line:ala:rexcludeSuperseededRec}\> Eliminate from $\mathit{pendConf}$ subsets of $\mathit{lastDec.conf}$\\
\nnll\label{line:ala:rcompleteMissedobjL}\> $\mathit{objL} := \mathit{objL}' \cup \mathit{objL}$\\
\nnll\label{line:ala:rcompleteMissedconfL}\> $\mathit{confL} := \mathit{confL}' \cup \mathit{confL}$\\
\\
\textbf{operation} \textit{extractLedger} $(\mathit{objL}', \mathit{confL}')$\\
\nnll\label{line:ala:rgetInputValue}\> \textbf{let} $\mathit{receivedValue} = (\sqcup [v \vert \exists p, \mathit{objL}'[p] = v], \sqcup [c \vert \exists p, \mathit{confL}'[p] = c])$\\
\nnll\label{line:ala:rledgerReturn}\>\> \textbf{return} $\mathit{receivedValue}$
\end{tabbing}
}
\vspace{-1.5mm}
 \hrule
\caption{Reconfigurable Accountable Lattice Agreement: Code for replica $r$.}
\label{fig:alapart3}
\end{figure}

\subparagraph*{Overview.} The clients propose values to the replicas which can either \emph{accept} them by issuing an ACK or \emph{reject} them by issuing a NACK. 
Once enough responses are gathered by the proposing client, it can accordingly either proceed to learn the value it proposed or to refine its proposal so it contains the missing information replicas raised. 
If no malicious replica tries to deviate, the values learnt are comparable and no accusations are raised. 
On the other hand, once a replica induces clients to learn incomparable values it is eventually detected and an accusation against it is produced.

The following definitions and boolean map are used in the algorithm and proofs of correctness:

\begin{definition}[$S$ satisfies configurations]
Let $S$ be a set of replicas, $\kappa$ a configuration, and $D$ a set of configurations.
We say that $S$ \emph{satisfies $D$ upon $\kappa$} iff, for each $d\subseteq D$, $S$ contains a majority of replicas in each configuration in the set $\kappa \cup {\cup d}$.
\end{definition}

\begin{definition}[Pending Configurations]
A configurations is called \emph{pending} as long as a client has received it but has not yet included it in the most recent decided configuration (lines~\ref{line:ala:excludeSuperseeded} and~\ref{line:ala:excludeSuperseededRec}). 
This set is comprised of the current client proposal, as well as configurations coming from ACKs (line \ref{line:ala:pending} and \ref{line:ala:includeACKPending}).
\end{definition}

The map $\textit{verify\_maj}: (\Sigma\times K \times 2^K)\to
\{\true,\false\}$ where $\mathit{verify\_maj(S, \kappa, D)=true}$ iff S satisfies $D$ upon $\kappa$. This map is used to indicate that the client gathered all the responses it needed.

\subparagraph*{Ledgers.}
Every client maintains a local \textit{ledger}, called \textit{ackL}, reserved to keep track of signed ACK messages the client received and their senders. 
Also, clients and replicas maintain two more ledgers to register the values
introduced in the system by their origin processes called \textit{objL} and \textit{confL}. By indexing a ledger $l$ by a process $p$ ($l[p]$), one can recover all the values signed by $p$ present in $l$.

\subparagraph*{Issuing a proposal.}
A client starts in a \textit{waiting} status and listens for values in its \textit{inBuffer} to include them
in a new proposal (lines \ref{line:ala:setobjL} and \ref{line:ala:setconfL}), not taking any values from the buffer while the executing a \textit{proposal}. 
Additionally, it must also listen for decisions made by other clients (lines \ref{line:ala:completeMissedobjL} and \ref{line:ala:completeMissedconfL}) including them in its proposal, preventing malicious replicas from keeping values from it.
It then proceeds to multicast its \textit{propV} to the replicas that might satisfy the pending configurations (variable \textit{dest}) it has seen upon the last decided configuration it came by (line \ref{line:ala:mcastProposal}) and waits for them to respond.

\subparagraph*{Treating Client Proposals.}
The replicas that receive the proposal extract the value from the ledger (line \ref{line:ala:getProposal}). This makes use of the operation \textit{extractLedger} which verifies that all the values came from existing clients, making these values valid. Each replica then checks whether the new proposal contains the join values it has already seen proposed (\textit{repV}), in which case they ack it (line \ref{line:ala:sendACK}) or not, in which case they nack it (line \ref{line:ala:nackknewValueSeen}), sending a complement to the ledger allowing the client to update its proposal. Benign replicas always forward their keys, destroying the old ones in the process, before responding to clients (line \ref{line:ala:serverforward}).

A replica cannot provide a client with outdated information because the timestamp used in the signature of its messages is only valid if it has been forwarded to the content it proposes and cannot be rolled back. Moreover, if a benign replica sees that a client isn't aware of a decision it has already come by, it will ignore the client proposal until it includes newer information (line \ref{line:ala:checklastDec}).

The function getMSG (line \ref{line:ala:convictSender})
takes a set of input values and returns the set of proposals or NACK messages that originally contained them.

\subparagraph*{Treating Replica Responses.}
Once the client gets an ACK from a replica, it includes the message in its ackL (line \ref{line:ala:appendACKList}) and registers the replica in its response set (line \ref{line:ala:appendACKRESPList}). Upon reception of a NACK, a client complements its \textit{objL} and \textit{confL} (lines \ref{line:ala:updateobjLProposal} and \ref{line:ala:updateconfLProposal}) and sets the NACK bool, including the replica in its response set (lines \ref{line:ala:incrementNACKCount} and \ref{line:ala:appendNACKRESPList}). When a client sees that it has gathered responses from a set of replicas that satisfies the pending configurations, it proceeds to check its NACK bool, as the presence of a NACK means that it cannot decide yet, and if it is false, then it will decide (line \ref{line:ala:checkNack}).

Each proposal gets a unique number (\textit{activePropNb}) so clients consider only reactions to the active proposal, ignoring late messages they might receive. Clients also ignore messages coming from replicas they already accused, as well as messages signed using timestamps that do not correspond to the configuration in their contents.

A client either waits until it gets responses from a set of replicas (keeping track via \textit{RESPSet}) that satisfies the pending configurations or until it gets a newer decided configuration from another client broadcast. It is necessary to get majorities in all those combinations of system configurations because the client doesn't know if any combination of them was learnt by another client and must be sure that it has reached all possible active configurations that can be learnt before it learns one by itself. Furthermore, the state transfer from one replica to another will be directly provided by this procedure, as once a client learns a configuration the object information is already in place, which is one of the advantages of this solution. It becomes then necessary to keep track of the state of the system by including information about which was the last combination of decisions seen (variable \textit{lastDec}), as well as pending configurations (variable \textit{pendConf}).

\subparagraph*{Issuing and Treating Convictions.}
We keep an array of all output values instead of just the current one,
as well as their corresponding ack ledgers, indexing the currently active entry by \textit{activeOutNb}.
This is necessary in order to monitor that after long delays in the network when two
correct clients re-establish their connection they can still check if in this period
their decisions were comparable (line \ref{line:ala:checkUncomparability}) and be able to
accuse processes that lead them to this incomparable state. The clients broadcast their accusations as well as their decisions. 

They avoid issuing redundant accusations by keeping track of the variations (line \ref{line:ala:checkDelta}). If a process gets new misbehavior proofs, it includes them on its accusations (line \ref{line:ala:addProof}).
\autoref{fig:verify-proof} shows one possible implementation of the \textit{verify\_proof} function,
checking that every issued accusation was made after a process tried to 
forge some signature, issued incoherent ACKs, tried to input values in the system in
discordance with the specification or decided something without gathering the necessary
acknowledgements.

Once a replica is accused by a client, the client begins to ignore the replica and the underlining application can for instance issue a reconfiguration effectively replacing it by one or more new replicas. The clients will then eventually learn comparable values once they join their values and the malicious replicas that try to subvert the system have been accused.

\FloatBarrier
\FloatBarrier
\subsection{Correctness}
We claim that the system of processes implemented following \autoref{fig:alapart1}, \autoref{fig:alapart2}, \autoref{fig:alapart3} solves RALA.
\begin{definition}
Let's define a state $s$ as being the value of the variable \textit{propV}. A state $s$ is considered as decided when the first client $c$ in state $s$ broadcasts its decision at line \ref{line:ala:bcastDecision}.
Moreover, we define \textit{s.lastDec} and \textit{s.pendConf} as being the value of these variables on the client $c$ at the moment of the state decision. Finally, \textit{s.confComb} is defined as the set $\{s.lastDec.conf \cup \{ \cup d\} \vert d \in 2^{s.pendConf}\}$
\end{definition}
\begin{definition}
We define then a graph $G_s$ whose vertices are the different decided states of the system plus the state $(\bot, initConf)$ and whose edges exist between two vertices $s$ and $s'$ whenever the following is true:
\[
    s \rightarrow s' \Leftrightarrow s \sqsubsetneq s' \land s.conf \in s'.confComb
\]
\end{definition}
\begin{lemma}
\label{lem:rala:increasingAccept}
At every benign replica, the variables \textit{repV}, \textit{pendConf} and \textit{lastDec} are monotonically increasing.
\begin{proof}
The variable \textit{repV} is updated in line~\ref{line:ala:updateAccepted} 
where it is assigned a new value which has passed the test in line \ref{line:ala:checkInclusion}, so the new value contains the old one. 

The other updates involving these variables in lines \ref{line:ala:updatelastDec}, \ref{line:ala:includePending} and \ref{line:ala:updateNAccepted} are joins where one of the operands is the old 
value, so the new values must contain the old ones.
\end{proof}
\end{lemma}
\begin{lemma}
\label{lem:rala:fork}
Given decided states $\overline s$, $s$ and $s'$ in $G_s$, if~$\overline s \rightarrow  s$, $\overline s \rightarrow s'$, $s'.\mathit{lastDec} \sqsubseteq s$, $s.\mathit{lastDec} \sqsubseteq s'$ then either there is an edge between $s$ and $s'$ or there is an accusation.
\begin{proof}
    From $\overline s \rightarrow  s$, $\overline s \rightarrow s'$ we derive that:
    \[
    \mathit{\overline s.conf \in  s.confComb \land \overline s.conf \in s'.confComb \implies~\overline s.conf \in s.confComb \cap s'.confComb}
    \]
    Because the decision only happens after triggering the event that begins in line \ref{line:ala:checkNack}, then it must be that the clients who decided these states got responses from replicas forming majority quorums in $\overline s.conf$ and they must therefore intersect in at least a replica $r$.
    
    Let us assume for now that $r$ followed the algorithm and behaved correctly. Let $c_s$ be the client that decided $s$ and $c_{s'}$ be the client that decided $s'$. Assuming w.l.o.g that the replica $r$ served the client $c_s$ before, using lemma \ref{lem:rala:increasingAccept} and observing that $s$ and $s'$ correspond to the first decision of these values, $s \sqsubsetneq s'$. Since $s'$ passed the test in line \ref{line:ala:checklastDec} in replica $r$, it means that $s.lastDec \sqsubseteq s'.lastDec$, moreover because we assume that $s'.\mathit{lastDec} \sqsubseteq s$ we can write:
    \[
    s.conf = \bigsqcup(\{s.lastDec.conf\} \cup s.pendConf) \sqsubseteq \bigsqcup(\{s'.lastDec.conf\} \cup s.pendConf) 
    \]
    \[
    \sqsubseteq \bigsqcup(\{s.conf\} \cup s.pendConf) = s.conf
    \]
    Furthermore, we see that all pending configurations in s which weren't included by the last decided configuration in $s'$ must also be pending in $s'$ because this information will be carried by the ack from replica $r$ (line \ref{line:ala:includeACKPending}). We can then conclude:
    \[
    s.conf = \bigsqcup(\{s'.lastDec.conf\} \cup s.pendConf)
    \]
    \[
    = \bigsqcup(\{s'.lastDec.conf\} \cup \{u \in s.pendConf, u \not \sqsubseteq s'.lastDec.conf\})
    \]
    \[
    \in \bigsqcup(\{s'.lastDec.conf\} \cup C \vert C \sqsubseteq s'.pendConf) = s'.confComb
    \]
    Hence there is an edge from $s$ to $s'$ in $G_s$ in this scenario.
    
    If the replica r didn't follow the algorithm and issued incomparable acks, then this event shall be detected and $r$ will be accused. $r$'s ack would be included in both clients $c_s$ and $c_{s'}$ \textit{ackLs} being broadcasted together with the decision in line \ref{line:ala:bcastDecision} and once the first client who decided and then received the other's decision go through the line \ref{line:ala:proofAcker}, it would find $r$ in both ledgers and accuse it. 
\end{proof}
\end{lemma}
\begin{lemma}
\label{lem:rala:merge}
All the infinite connected components of $G_s$ have the same suffix.
\begin{proof}
    Because we assume that the system reconfigures finitely many times, there is a point where all the decisions regarding the different configurations the system passed, which were broadcasted in line \ref{line:ala:bcastDecision}, arrive at the recipient clients. They'll process the configuration included in these values in lines \ref{line:ala:updateConfRec} and \ref{line:ala:completeMissedconfL} and the use of the forward secure signatures will prevent them from processing messages of old replicas. As a result, all the clients will have the same values of \textit{lastDec.conf} and \textit{propV.conf}, meaning that they will contact the same replicas and need to form a majority in the active configuration $lastDec.conf \sqcup  propV.conf$, where their majority quorums intersect. As seen in the lemma \ref{lem:rala:fork} if any malicious replica tries to issue incomparable ACKs the clients will accuse it and ignore it from this point onward. They will retry the proposal again until none of the replicas in the majority misbehaves. 
    
    Let $s$ be the first state decided in this scenario, henceforth the states will be totally ordered, sharing the same configuration which is always present in confComb, meaning that these states are connected. The graph will only have one growing branch and any new state $s'$ in it will be a descendant of $s$. Finally, this branch will be infinite because the clients never stop proposing.
\end{proof}
\end{lemma}
\begin{theorem}
\label{theo:rala:validity}
The system of processes implemented following \autoref{fig:alapart1}, \autoref{fig:alapart2}, \autoref{fig:alapart3} provides \emph{validity}.
\begin{proof}
    The learnt values by a client are extracted from its \textit{objL} and \textit{confL}.
    
    First of all, at the beginning of a proposal (lines \ref{line:ala:setobjL} and \ref{line:ala:setconfL}) the value present in the input buffer is read and put into the ledgers, guaranteeing that when a decision is made it shall be present in it.
    
    These variables are then modified in lines \ref{line:ala:updateobjLProposal}, \ref{line:ala:updateconfLProposal},
    \ref{line:ala:completeMissedobjL} and
    \ref{line:ala:completeMissedconfL}. Therefore, the values included into them
    either come from the the input buffer from the clients where they are signed, or they are informed by replicas nacking proposals after passing signature check
    or by the information of other clients decisions. We conclude that the values learnt always come from the client input buffers directly or indirectly. 
\end{proof}
\end{theorem}
\begin{theorem}
\label{theo:ala:consistency}
The system of processes implemented following \autoref{fig:alapart1}, \autoref{fig:alapart2}, \autoref{fig:alapart3} provides \emph{completeness}.
\begin{proof}
   By Lemma \ref{lem:rala:fork} whenever the graph $G_s$ forks an accusation is issued. Each fork occur when clients learn incomparable values and are caused by some Byzantine replicas which are eventually accused. Moreover, by lemma \ref{lem:rala:merge} the system cannot be indefinetely forked and all inconsistencies are eventually solved when no new accusations are issued as required.
\end{proof}
\end{theorem}
\begin{theorem}
\label{theo:ala:convictionstab}
The system of processes implemented following \autoref{fig:alapart1}, \autoref{fig:alapart2}, \autoref{fig:alapart3} provides \emph{accusation stability}.
\begin{proof}
    The set of accused processes is reflected in the algorithm via the variable
    \textit{accusation} which is updated in lines \ref{line:ala:proofAcker}, \ref{line:ala:convictSender},
    and \ref{line:ala:addProof}. As one can see, they either attribute this variable
    to a union where one of the operands is itself or a value is explicitly included into it.
    Therefore after each update the new value must, by the definition of these operations,
    include the old one, i.e. the accusation set is monotonically increasing.
\end{proof}
\end{theorem}
\begin{theorem}
\label{theo:ala:accuracy}
The system of processes implemented following \autoref{fig:alapart1}, \autoref{fig:alapart2}, \autoref{fig:alapart3} provides \emph{accuracy}.
\begin{proof}
An accusation can be issued in lines \ref{line:ala:noPendingConv}, \ref{line:ala:proofAcker} and \ref{line:ala:convictSender}.

The first occurrence checks that an ACK was issued but the matching configuration proposal wasn't added by the replying replica as described in line \ref{line:ala:includePending} meaning that this line was skipped.

On the second case the decision of incomparable values requires, as seen earlier in \autoref{theo:ala:consistency} 
that a replica acknowledged incomparable values, violating the behavior of benign replicas
described by Lemma \ref{lem:rala:increasingAccept}. Having two ACKs signed by the same process
for incomparable values characterises an irrefutable proof and the replicas in it will be a non-empty subset 
of $M$ because they deviated from the algorithm.

On the last case a replica will be caught providing fake signatures, which is by itself enough to accuse it as this is a clear deviation from the algorithm.
\end{proof}
\end{theorem}
\begin{theorem}
\label{theo:ala:authenticity}
The system of processes implemented following \autoref{fig:alapart1}, \autoref{fig:alapart2}, \autoref{fig:alapart3} provides \emph{authenticity}.
\begin{proof}
Authenticity follows from our cryptographic assumptions and specially from three
properties of the underlying system:
    \begin{itemize}
        \item Every message contains a signature; 
        \item The signatures can be verified by a public function;
        \item No other process can sign on behalf of a correct process.
    \end{itemize}
\end{proof}
\end{theorem}
\begin{theorem}
\label{theo:ala:agreement}
The system of processes implemented following \autoref{fig:alapart1}, \autoref{fig:alapart2}, \autoref{fig:alapart3} provides \emph{agreement}.
\begin{proof}
    Agreement is a direct consequence of the dissemination of information implemented
    in the algorithm. Every accusation is broadcasted and every message containing an accusation
    is analysed and, if it holds, leads to the adoption of the information (line \ref{line:ala:newConvictionFound}).
\end{proof}
\end{theorem}
\begin{theorem}
\label{theo:ala:liveness}
The system of processes implemented following \autoref{fig:alapart1}, \autoref{fig:alapart2}, \autoref{fig:alapart3} is \emph{alive}.
\begin{proof}
    After the system stops reconfiguring a client can eventually receive the last configuration learnt by the broadcast in line \ref{line:ala:bcastDecision} and then every client will contact active configurations. If a client then starts a proposal when receiving a value $v$ in its input buffer, a majority of replicas in all active configurations shall eventually respond to this client, following our majority of correct replicas assumption in this scenario. From this point on, all decisions shall contain $v$ as the last learnt configuration all clients contact will provide a majority of replicas that include $v$. 
\end{proof}
\end{theorem}

\section{Related Work}
\label{sec:related}
%

\subparagraph*{Accountability.}
In security terms, accountability ensures that the actions of an entity can be traced solely to that entity. This supports non-repudiation, deterrence, fault detection, and after-action recovery. Distributed computing research has focused for many years on failure detection ~\citep{CHT96,DFGHKT04,KMM03}, a  close relative of accountability. By identifying faulty processes, failure detection helps the distributed computation to make progress in a safe way, but does not provide evidences of misbehaviors that can be verified by a third party. 
To the best of our knowledge, PeerReview~\citep{peerreview} was the first  proposing a general solution  to provide accountability as an add-on feature for any distributed protocol. In PeerReview each process in the system records messages in tamper-evident logs: an auditor can challenge a process, retrieve its logs, and simulate the original protocol to ensure that the process behaved correctly. By doing so, any observable deviating action can be traced back to at least one Byzantine process that was responsible for it. The main issue is that for an auditor to prove that a process is Byzantine it must receive a response to the challenge from the process. If no response is received, the auditor cannot determine whether the process is faulty or not. As a result, some Byzantine processes might  be suspected forever and never proven guilty. This limitation is common to distributed protocols that are not designed to provide accountability.

Polygraph~\citep{civit2021polygraph} equips  Byzantine Consensus with an accountability mechanism. 
%
%
As in our system, the very messages sent during the protocol execution carry the necessary information to construct a proof in case of Consensus agreement violation. This way, there is no need to query processes to collect evidences and construct a proof. 
Fairledger~\cite{fairledger} and LLB (Long-Lived Blockchain)~\cite{llb} are consensus-based state-machine replication protocols that are able to detect consistency violations in consensus instances and reconfigure themselves.
%
In contrast, we do not rely on consensus for reconfiguration and propose a purely asynchronous accountable and reconfigurable service.  

\subparagraph*{Lattice agreement.}
Attiya et al. ~\citep{lattice-hagit} introduced the (one-shot) lattice agreement abstraction and, in the shared-memory context, described a wait-free reduction of lattice agreement to atomic snapshot. Falerio et al. ~\citep{gla}
introduced the long-lived version of lattice
agreement (adopted in this paper), called generalized lattice agreement, and described an asynchronous
message-passing implementation of lattice agreement assuming a majority
of correct processes. 
In the Byzantine failure model,  Di Luna et al ~\citep{bgla} proposed for the first time a solution for Byzantine asynchronous generalized lattice agreement, later improved by ~\citep{bla}. All these algorithms propose a fault-tolerant approach where safety and liveness are guaranteed with $f<n/3$ Byzantine processes and authenticated channels. In our accountability approach, liveness and recovery from safety violations are guaranteed with $f<n/2$ Byzantine processes and authenticated channels. 

\subparagraph*{Asynchronous reconfiguration.}
Dynastore~\citep{dynastore} 
was the first solution emulating a
reconfigurable atomic read/write register without consensus:
clients can asynchronously propose  incremental additions or removals
to the system configuration. Since proposals commute, concurrent proposals are collected together
without the need of deciding on a total order. In~\citep{smartmerge} it has been observed that asynchronous reconfiguration can be handled using an external reliable lattice-agreement object. 
Reconfigurable lattice agreement~\cite{rla} enables   reconfigurable versions of a large class of
objects and abstractions, including state-based CRDTs~\cite{crdt}, atomic-snapshot, max-register,
conflict detector and commit-adopt.

In the Byzantine fault model, Dynamic Byzantine storage~\cite{alvisi2000dynamic,martin2004framework} allows a trusted
\emph{administrator} to issue ordered reconfiguration
calls that might also change the set of replicas.
More recently, \citep{rbla}~describes a generic Byzantine fault-tolerant reconfigurable lattice agreement, implemented without assuming a trusted administrator. 

The reconfiguration technique used in this paper takes inspiration from~\citep{rla} while been enriched with the use of forward-secure signatures as proposed in~\citep{rbla} to protect the system from Byzantine replicas belonging to old configurations. 
Note that none of the cited work provide proof-of-misbehavior of Byzantine processes.

\FloatBarrier

\section{Concluding remarks}
\label{sec:discussion}
In this paper, we propose the first design of an asynchronous replicated system that not only detects misbehavior that affects its safety properties, but is also able to mitigate misbehaving replicas by reconfiguration.  
Compare to earlier~\cite{rambo,fairledger} and concurrent~\cite{llb} work, we do not employ consensus to agree on the evolving configurations.    
The algorithm described in this paper can be improved and generalized in multiple ways. Below we discuss some of them. 

\subparagraph*{Garbage collection.} 
%
In the current version of our algorithm, every process locally maintains a complete history of updates, and periodic reinitialization of the system is an important issue.   
In particular, it appears challenging to reinitialize the set of accusations, as a slow client may never be able to be convinced  that a compromised replica is not trustful anymore.  
One may think, e.g., of a periodic instances of a consensus protocol among the clients to agree on the new initial system state, running in parallel with our algorithm.
Altogether, periodic ``truncation'' of the ever-growing state in an asynchronous protocol remains an interesting question for the future work.  
\ignore{
 The reason why comes from the fact that if a replica in a given moment manages to completely split the clients in two configurations, then the only way to correctly accuse it is by receiving the broadcasts of the incomparable decisions and checking against the acks received at the time the clients learnt incomparable values. This could be solved by sporadically running in parallel with the system a garbage collection protocol which also synchronises the clients allowing to safely removing old information from the ledgers.
}

\subparagraph*{Complexity.} 
Similar to earlier solutions of (generalized) lattice agreement~\cite{gla,rla}, the latency of learning a value in our algorithm (in the number of asynchronous query-response rounds, assuming that the configuration does not change) is proportional to the number of concurrently proposed values. 
It remains unclear if there is an asymptotically faster algorithm. 
There are interesting solutions for \emph{one-shot} Byzantine lattice agreement that take $\log k$ rounds for $k$ proposed values~\cite{bla-garg}, but we do not have a comparable long-lived implementation.  

For simplicity, in our algorithm, the sizes of messages grow linearly with the number of distinct values learnt by the clients.
One can improve this by sending relative updates instead of complete histories in PROPOSE and DECISION messages. 
The size of ACK and NACK messages already grow much slower, as they use digests of corresponding proposals and only contain information about changes:
in the case of ACKs, these changes consist of the pending configurations since last decision and in the case of NACKs---with respect to the proposed value the replica is responding to. 
An ACCUSATION message has asymptotic complexity of an ACK message. 
%
The issue of maintaining bounds on ever-growing message sizes is related to the more general question of garbage-collection and reinitialization.
\ignore{
ACK messages and NACK messages also grow linearly with the number of distinct proposed values in the system for the same reason and are issued each time a proposal reaches a replica. 
As for DECISION messages, they have a part which is once more linear with the number of distinct values in the system, but also a part which grows linearly with the number of replicas, corresponding to the ACK ledger. As for ACCUSATIONS, they have variable size, but are most of the time consisted of a couple of messages issued by a replica. For this reason if we assume that the system is comprised of $c$ clients, $r$ replicas and has exchanged $n$ different values, the message size complexity will be $O(cnr)$ and the number of messages exchanged will also be $O(cnr)$.
}

\subparagraph*{Clients: Byzantine and heavy.}

Early proposals of quorum-based fault-tolerant storage systems typically assumed that clients are benign (see, e.g., \citep{MAD02-storage}). 
While the effect of Byzantine \emph{writers} can be mitigated using erasure coding~\citep{GWGR04-storage} or voting~\citep{LR06-storage}, it appears nontrivial to handle malicious reconfiguration requests. 
Indeed, a Byzantine client can block progress by plunging the system in constant reconfiguration, or break safety of the replicated data by rendering the system to a compromised configuration.   
How to handle such attacks is an intriguing challenge.

Assuming that the clients are benign enables assigning them with a major part of the total work. 
%
%
This  results in linear message complexity: the replicas only passively respond to clients' queries.

Alternatively, we may follow earlier work on asynchronous Byzantine reconfiguration~\cite{rbla}, and assume an external \emph{access control} mechanism ensuring that inputs from the clients (including reconfiguration calls) are ``acceptable''.
In particular, the proposed configurations should satisfy the configuration availability condition (Section~\ref{sec:rala}): every combination of candidate configurations should contain enough correct replicas.  
Also, the access control mechanism should provide a verification procedure that would allow the third party to verify validity of reconfiguration requests. 
The clients are then only responsible for submitting valid to the set of replicas.
The resulting algorithm will, however, likely to be more costly in terms of message complexity, as each reconfiguration will have to handle each of the valid requests.  

Our algorithm can also be easily extended to accommodate partitions of the clients into (benign) administrators and (Byzantine-prone) users, along the lines of~\cite{martin2004framework,fairledger}.  
%
%

For completeness, in Appendix~\ref{sec:A1LA}, we describe a specification and a corresponding implementation of a \emph{one-shot} lattice agreement abstraction that assumes that both clients and replicas can be Byzantine. 
Our system is particularly well suited for the client-administrator approach as the reconfiguration requests are issued by the proposing entities (in this case an administrator) and not the entities maintaining the system (the replicas).

\subparagraph*{Reconfiguration strategy.}
In this paper, we delegated the task of choosing new configurations to the clients.  
The clients are free to reconfigure the system even if no new misbehaving replicas are detected. 
The only requirement we impose on the configurations proposed by the clients is that resulting configurations must remain available (Section~\ref{sec:rala}). 
But one may think of more explicit reconfiguration strategies. For example, each time a new misbehaving replica is detected, it is replaced with a new one taken from a ``pool'' of correct replicas (a similar approach is proposed in LLB~\cite{llb} for consensus-based reconfiguration).   


\FloatBarrier

\bibliographystyle{abbrv}
\bibliography{references}  


\appendix
\section{Accountable Lattice Agreement with Byzantine Clients} 
\label{sec:A1LA}
%
In this section, we discuss a \emph{one-shot} static version of accountable lattice agreement that considers that both clients and replicas might be Byzantine. 
 
\subsection{Problem statement.}
The \emph{general accountable one-shot lattice agreement} (\emph{A1LA}) abstraction, defined on a lattice
$(\Lat, \sqsubseteq)$, takes, as a single input, an element in $\Lat$ and produces, as an output,
a pair of an element in $\Lat$ 
and a set of \emph{accusations}.
Again, an accusation is a pair $(A,P)$ where $A \subset \Pi$ and $P$ is a \emph{proof of misbehavior}.
And we assume that the proof can be independently verified by a third party
through a boolean map $\textit{verify-proof}: (2^{\Pi}\times \cP)\to \{\true,\false\}$. 

We say that this version is general because both clients and replicas can be malicious. The system contains N replicas where a majority of them are correct.
Let $U\subseteq B$ be the finite set of benign clients that proposed values in that run, and let  $u_i$ denote the value proposed by a process $p_i\in U$.
Let $V_B$ and $V_C$ be the sets of, respectively, benign and correct clients that learned values in that run, and let $v_i$ denote the value learned by a process $p_i\in V_B$
(obviously, $V_C\subseteq V_B \subseteq U$). 
The A1LA abstraction satisfies the following properties:

\begin{itemize}
\item {\bf Validity.} The value learnt by a benign client $c_i$ with input value $u_i$ is a join of values 
    proposed by clients in $U$ (including $c_i$), at most $\lvert M \rvert$ values coming from
    $M$, and $u_i$:
    \[
    \forall c_i \in V_B: u_i\sqsubseteq v_i \wedge v_i \sqsubseteq \sqcup(\{u_j | c_j \in U\}\cup F),\; F\subseteq \Lat,\; |F|\leq |M|.
    \]
\item {\bf Consistency.} Either the values learned by the correct clients are totally ordered: 
\[
 \forall c_i, c_j \in V_C: v_i\sqsubseteq v_j \vee v_i\sqsubseteq v_j.
\]
or every correct process eventually accuses a set of processes.

\item {\bf Accuracy.} If a benign process $p_i$ accuses $A$ (with $P$), then $A$ is a
    subset of $M$ and $P$ contains a proof against each process in $A$.
    \[
        \forall p_i \in B, A \subseteq M \wedge
        \textit{verify-proof}(A,P)
    \]
\item {\bf Authenticity.} It is computationally infeasible to construct $A\cap B\neq \emptyset$ and $P\in\cP$ such that $\textit{verify-proof}(A,P)=\true$.       
  
\item {\bf Liveness.} If a correct client proposes a value, it eventually learns a value or accuses a set of processes.
  
\end {itemize}  
				
One can see that a benign client can accuse a set of processes only if there is at
least one Malicious process ($M\neq \emptyset$) and in the absence of malicious processes
no proofs of misbehavior are created.
The Consistency property guarantees that either \emph{correct} clients learn comparable values or some malicious process will be accused. 
Notice that  we cannot avoid executions in which a \emph{Benign} but not correct client learns a value that is inconsistent with a value learnt by another benign client if there are malicious processes without issuing accusations. 

\subsection{The algorithm}
Our solution to A1LA is presented in \autoref{fig:a1laclient1}, \autoref{fig:a1laclient2} and \autoref{fig:a1laserver}. As before, each method is executed in its entirety without being interrupted.
Each client might be in an \textit{active} state where it proposes values and takes steps towards
learning something new, re-proposing if necessary or in a \textit{passive} state where
it only reacts to other client proposals.
On start-up clients that receive input values from the application sign them and put them to their
input ledgers (line \ref{line:a1la:setLedger}) and proceed
to propose it to the system by multicasting (\ref{line:a1la:bcastProposal}).

When a replica receives a proposal with a ledger, it extracts the proposed value by the
other process merging the new inputs to its own ledger (\ref{line:a1la:updateLedgerProposal}).
Before treating the ledger a verification is made to guarantee its integrity (\ref{line:a1la:checkByzantineSender}). Once the
message is validated, there might be inconsistencies in the ledger introduced by malicious
processes that tried to insert more than one value in the system and an accusation might be issued
(\ref{line:a1la:checkDoubleProposal}). Otherwise the ledger is consistent and an ACK
can then be produced if the proposal comprises the previously received ones (\ref{line:a1la:sendACK}),
otherwise a NACK shall be sent informing the proposer of values that it didn't include in its proposal
via a complement to its ledger (\ref{line:a1la:nacknewValueSeen}).

Received ACKs are discarded if they correspond to old proposals, or if they come from a process
whose ACK has already been accounted in the current proposal, or if they don't match
the proposed value they were sent for. Note that the latter is in itself a sign of byzantine
behavior but doesn't constitute an irrefutable proof of misbehavior as the conflicting
messages, i.e. the proposal and the ACK, are signed by different processes. When the ACK
is valid it is included in the \textit{ackL} (\ref{line:a1la:appendACKList})
and the respective counter is incremented (\ref{line:a1la:incrementackCnt}).
Similarly, outdated NACKs are ignored when received,
as well as empty ledgers, ledgers who don't include new values (\ref{line:a1la:nonDelta}).
Once the ledger is validated, the proposed value is set to include the information
patch (\ref{line:a1la:updateProposal}) and the counter of NACKs is incremented
(\ref{line:a1la:incrementNackCnt}).

After a client has received responses from at least a majority of replicas (\ref{line:a1la:ackCntCheck}
and \ref{line:a1la:nackCntCheck}) for its proposal it proceeds to evaluate if it
can decide on a value or not, in case it has not accused any malicious processes
along the way. If a NACK has been received, it tries 
to learn its new proposal (\ref{line:a1la:checkNack}) which includes the missing
values in the previous attempt, otherwise it decides upon its proposal and broadcasts
it alongside the ledger of ACKs it has collected (\ref{line:a1la:bcastDecision}).
At this point clients who are proposing incomparable values to the one decided check
the ACKs they received so far as byzantine processes can lead the system to decide 
incomparable values by issuing contradictory ACKs for different processes, which can
be detected at this point and lead to their accusation (\ref{line:a1la:pickByzantineAcker}).

\begin{figure}[!htb]
\hrule \vspace{1mm}
 {\small
\begin{tabbing}
 bbbb\=bbbb\=bbbb\=bbbb\=bbbb\=bbbb\=bbbb\=bbbb \=  \kill
\textbf{Local variables:} \\
\textit{status}, initially passive  \commentline{Boolean for the current state: passive or active }\\
\textit{ackCnt}, initially 0  \commentline{The number of \emph{acks} received for the active proposal}\\
\textit{nackCnt}, initially 0  \commentline{The number of \emph{nacks} received for the active proposal}\\
\textit{activePropNb}, initially $-1$ \commentline{The number of \emph{nacks} received for the active proposal} \\
\textit{propV}, initially $\bot$ \commentline{The value being proposed}\\
\textit{inL}, initially empty \{ KV table (ledger) init. w/ proposed values signed by their originators \} \\
\textit{ackL}, initially empty \commentline{Key value table holding received signed acks by replicas} \\
\textbf{Input:}\\
\textit{initV} \commentline{Value initially proposed by the process, provided by external source} \\
\textbf{Outputs:}\\
\textit{outV}, initially $\bot$ \commentline{Value learnt by the client}\\
\textit{accusation}, initially $\emptyset$ \commentline{Proofs of misbehavior gathered}\\
\\
\textbf{upon} \textit{startup} \textbf{if} $\mathit{initV} \neq \bot$\\
\nnll\label{line:a1la:setProposal}\> $\mathit{propV} := \mathit{initV}$\\
\nnll\label{line:a1la:setLedger}\> include \textit{signed} $\mathit{initV}$ to $\mathit{inL}$\\
\nnll\label{line:a1la:callPropose}\> Propose\\
\\
\textbf{operation} \textit{Propose} \\
\nnll\label{line:a1la:activation}\> $\mathit{status} := \mathit{active}$\\
\nnll\label{line:a1la:updateProposal}\> $\mathit{propV} := \mathit{extractLedger(inL, c)}$\\
\nnll\label{line:a1la:incrementProposal}\> $\mathit{activePropNb} := 
\mathit{activePropNb} + 1$\\
\nnll\label{line:a1la:scrapAcks}\> clear \textit{ackL} \\
\nnll\label{line:a1la:counterReset}\> $\mathit{ackCnt} := \mathit{nackCnt} := 0$\\
\nnll\label{line:a1la:bcastProposal}\> \textit{multicast} $\langle \textit{PROPOSAL}, 
    (inL, activePropNb) \rangle$ to \textit{Servers}\\
\\
\textbf{operation} \textit{EvaluateDecision} \\
\nnll\label{line:a1la:checkNack}\> \textbf{if} $\mathit{nackCnt} > 0$ \textbf{then} Propose \\
\nnll\label{line:a1la:decide}\> \textbf{else} \\
\nnll\label{line:a1la:updateOutput}\>\> $\mathit{outV} := \mathit{propV}$ \\
\nnll\label{line:a1la:decidePassive}\>\> $\mathit{status} := \mathit{passive}$ \\
\nnll\label{line:a1la:bcastDecision}\>\> \textit{multicast} $\langle \mathit{DECISION}, 
(\mathit{outV}, \mathit{ackL}) \rangle$ to \textit{Servers}\\
\\
 \textbf{upon} \textit{receive} $\langle \textit{ACK}, (propV, \mathit{activePropNb}) \rangle$ \\
\>from given replica $r$ AND $r \not \in ackL$ $\mathit{status} = \mathit{active}$\\
\nnll\label{line:a1la:appendACKList}\> append $q$'s ack to $\mathit{ackL}$\\
\nnll\label{line:a1la:incrementackCnt}\> $\mathit{ackCnt} := \mathit{ackCnt} + 1$ \\
\nnll\label{line:a1la:ackCntCheck}\> \textbf{if} $\mathit{ackCnt} + \mathit{nackCnt} \geq \lceil \frac{N+1}{2} \rceil$ \textbf{then} EvaluateDecision\\
\\
\textbf{upon} \textit{receive} $\langle \textit{NACK}, (\mathit{\Delta Ledger}, \mathit{activePropNb}) \rangle$ from process $q$\\
\> AND $\mathit{status} = \mathit{active}$ AND $\mathit{\Delta Ledger} \neq \emptyset$ \\
\nnll\label{line:a1la:getDeltaValue}\> $\mathit{\Delta Value} = \mathit{extractLedger}(\mathit{\Delta Ledger}, q)$\\
\nnll\label{line:a1la:nonDelta}\label{line:a1la:nonDeltaReturn}\> \textbf{if} $\mathit{\Delta Value} \sqsubseteq \mathit{propV}$ \textbf{return}\\
\nnll\label{line:a1la:updateLedgerAfterNACK}\> $\mathit{inL} := \mathit{inL} \cup (\mathit{inL}')$\\
\nnll\label{line:a1la:incrementNackCnt}\> $\mathit{nackCnt} := \mathit{nackCnt} + 1$ \\
\nnll\label{line:a1la:nackCntCheck}\> \textbf{if} $\mathit{ackCnt} + \mathit{nackCnt} \geq \lceil \frac{N+1}{2} \rceil$ \textbf{then} EvaluateDecision
\end{tabbing}
}
\vspace{-1.5mm}
 \hrule
\caption{Accountable One-Shot Lattice Agreement: Code for client $c$ part 1.}
\label{fig:a1laclient1}
\end{figure}
\begin{figure}[!htb]
\hrule \vspace{1mm}
 {\small
\begin{tabbing}
 bbbb\=bbbb\=bbbb\=bbbb\=bbbb\=bbbb\=bbbb\=bbbb \=  \kill
\textbf{upon} \textit{receive} $\langle \textit{DECISION}, (\mathit{outV}', \mathit{ackL}') \rangle$ from process $q$\\
\nnll\label{line:a1la:fakeDecision}\> \textbf{if} $\exists (p, v) \in \mathit{ackL'}
\vert v \neq \mathit{outV'}$\\
\nnll\label{line:a1la:convictDecider}\>\> $\mathit{accusation} := \mathit{accusation}
\cup \{(\mathit{q}, \mathit{DECISION})\}$\\
\nnll\label{line:a1la:passiveWrongDecision}\>\> $\mathit{status} = \mathit{passive}$\\
\nnll\label{line:a1la:ignoreDecision}\>\> \textbf{return} \\
\nnll\label{line:a1la:checkUncomparability}\> \textbf{if} $\mathit{outV}' \not \sqsubseteq
\mathit{propV} ~\&\&~ \mathit{propV} \not \sqsubseteq \mathit{outV}'$ \textbf{then}\\
\nnll\label{line:a1la:pickByzantineAcker}\>\> \textbf{let} $M = \{m \vert m \in \mathit{ackL} ~\&\&~ \mathit{ackL}'\}$ \textbf{do}\\
\nnll\label{line:a1la:proofAcker}\>\>\> \textbf{foreach} $m \in M$ \textbf{do} include
$(b, \{\mathit{ackL}[b], \mathit{ackL}'[b]\})$ to \textit{accusation}\\
\nnll\label{line:a1la:byzantineAckerPassive}\>\>\> \textbf{if} $\lvert M \rvert > 0$ \textbf{then} $\mathit{status} := \mathit{passive}$\\
\\
\textbf{operation} \textit{extractLedger} $(\mathit{inL}', \mathit{sender})$\\
\nnll\label{line:a1la:checkByzantineSender}\> \textbf{if} $\exists~\mathit{process}~p \in \mathit{inL}'$ with invalid signature \textbf{then}\\
\nnll\label{line:a1la:convictSender}\>\> $\mathit{accusation} := \mathit{accusation}
\cup \{(\mathit{sender}, \mathit{getPropNACKMSG(inL')}\}$\\
\nnll\label{line:a1la:convictSSenderpassive}\>\> $\mathit{status} := \mathit{passive}$\\
\nnll\label{line:a1la:convictSSenderreturn}\>\> \textbf{return} $\emptyset$\\
\nnll\label{line:a1la:updateLedgerProposal}\> $\mathit{inL''} := \mathit{inL} \cup (\mathit{inL}')$\\
\nnll\label{line:a1la:checkDoubleProposal}\> \textbf{let} $M = \{m \vert m \in \mathit{inL''} ~\&\&~ \lvert \mathit{inL''}[m] \rvert > 1\}$ \textbf{do}\\
\nnll\label{line:a1la:proofOrigin}\>\> \textbf{foreach} $m \in M$ \textbf{do} include $(m, \mathit{getPropNACKMSG}(\mathit{inL''}[m]))$ to \textit{accusation}\\
\nnll\label{line:a1la:convictSourcepassive}\>\> $\mathit{status} := \mathit{passive}$\\
\nnll\label{line:a1la:returnConvictSourcepassive}\>\> \textbf{return} $\emptyset$\\
\nnll\label{line:a1la:getInputValue}\> \textbf{let} $\mathit{receivedValue} = \sqcup [v \vert \exists p, \mathit{inL}'[p] = v]$\\
\nnll\label{line:a1la:ledgerReturn}\>\> \textbf{return} $\mathit{receivedValue}$
\end{tabbing}
}
\vspace{-1.5mm}
 \hrule
\caption{Accountable One-Shot Lattice Agreement: Code for client $c$ part 2.}
\label{fig:a1laclient2}
\hrule \vspace{1mm}
 {\small
\begin{tabbing}
 bbbb\=bbbb\=bbbb\=bbbb\=bbbb\=bbbb\=bbbb\=bbbb \=  \kill
 \textbf{Local variables:} \\
\textit{inL}, initially empty\\
\commentline{Key value table holding initially proposed values signed by their originators} \\
\textit{repV} initially $\bot$ \commentline{Value held by the replica}\\
\textit{accusation}, initially $\emptyset$ \commentline{Proofs of misbehavior gathered by the replica}\\
\\
\textbf{upon} \textit{receive} $\langle \textit{PROPOSAL}, (\mathit{inL}', \mathit{activePropNb}') \rangle$ from process $q$\\
\nnll\label{line:a1la:getProposal}\> $\mathit{propV'} := \mathit{extractLedger}(\mathit{inL}')$\\
\nnll\label{line:a1la:checkInclusion}\> \textbf{if} $\mathit{repV} \sqsubseteq \mathit{propV}'$ \textbf{then}\\
\nnll\label{line:a1la:sendACK}\>\> send $\langle \textit{ACK}, (propV', \mathit{activePropNb}') \rangle$ to $q$\\
\nnll\label{line:a1la:updateAccepted}\>\> $\mathit{repV} := \mathit{propV}'$\\
\nnll\label{line:a1la:newValueSeen}\> \textbf{else} \\
\nnll\label{line:a1la:updateNAccepted}\>\> $\mathit{repV} := \mathit{repV} \sqcup \mathit{propV}'$\\
\nnll\label{line:a1la:nacknewValueSeen}\>\> send $\langle \textit{NACK}, (\mathit{inL} - \mathit{inL}', \mathit{activePropNb}') \rangle$ to $q$\\
\\
\textbf{operation} \textit{extractLedger} $(\mathit{inL}', \mathit{sender})$\\
\{ Identitical to client operation with the same name without lines \ref{line:a1la:convictSSenderpassive} and \ref{line:a1la:convictSourcepassive} \}
\end{tabbing}
}
\vspace{-1.5mm}
 \hrule
\caption{Accountable One-Shot Lattice Agreement: Code for replica $r$.}
\label{fig:a1laserver}
\end{figure}

\subsection{Correctness}
We claim that the algorithm presented in \autoref{fig:a1laclient1}, \autoref{fig:a1laclient2} and \autoref{fig:a1laserver}
solves the General Accountable One-Shot Lattice Agreement.
\begin{lemma}
\label{lem:increasingProposal}
At every benign process, the variable \textit{propV} is monotonically increasing.
\begin{proof}
The variable is updated only in line \ref{line:a1la:updateProposal}.
As one can see, it is a join operation where one of the operands is the previous value, hence the
new value by definition contains the old value.
\end{proof}
\end{lemma}
\begin{lemma}
\label{lem:onePerProcess}
If a benign process $p$ learns a value $v$, then $v$ cannot contain two or more values
signed by the same process.
\begin{proof}
A process adds values to its proposal $\textit{propV}$, when it receives a  \emph{nack} response. 
The insertion is then subject
to verification following line \ref{line:a1la:checkDoubleProposal}. 
Process $p$
will only proceed to a deciding a value if the list comprehension 
yields an empty list, in which case there is at most one value introduced on its proposal
per process in the system.
\end{proof}
\end{lemma}
\begin{lemma}
\label{lem:increasingAccept}
At every benign process, the variable \textit{repV} is monotonically increasing.
\begin{proof}
The variable is updated in lines~\ref{line:a1la:updateAccepted} 
and~\ref{line:a1la:updateNAccepted}.
The first update assigns to this variable a new value which has passed the test in line \ref{line:a1la:checkInclusion}, so the new value contains the old one. 
Similarly to \textit{propV}, the second update is a join where one of the operands is the old 
value, so the new value must contain the old one.
\end{proof}
\end{lemma}
\begin{theorem}
\label{theo:consistency}
The Algorithm presented in \autoref{fig:a1laclient1}, \autoref{fig:a1laclient2} and \autoref{fig:a1laserver}
provides \emph{consistency}.
\begin{proof}
Suppose, by contradiction, that two benign processes $p$ and $q$ learned two incomparable values $v'$ and $v''$.
The majority that acknowledged $v'$ at $p$ must intersect with the majority that acknowledged $v''$ at $q$.
Let $r$ be any process in the intersection. 
If $r$ is not Byzantine then by Lemma~\ref{lem:increasingAccept}, $v'$ and $v''$ must
be comparable and \emph{consistency} will hold.

Otherwise, $r$ must have acked incomparable values, which shall be detected in line 
\ref{line:a1la:pickByzantineAcker} meaning that the processes that output incomparable
values will accuse $r$.

\end{proof}
\end{theorem}
\begin{theorem}
\label{theo:validity}
The Algorithm presented in \autoref{fig:a1laclient1}, \autoref{fig:a1laclient2} and \autoref{fig:a1laserver}
provides \emph{validity}.
\begin{proof}
The inclusion of the process own proposal follows from Lemma \ref{lem:increasingProposal}
with the initialisation of \textit{propV} to \textit{initV}, remarking that \textit{outV} is
but one of the values taken by \textit{propV}.

As for the cap on the number of values coming from byzantine processes, suppose that there
are at least $\lvert M \rvert + c$, where $c \in \mathbb{N}^*$, values coming from byzantine
processes. It means that at least one byzantine process $b$ signed two or more initial
values that are output by a benign process. Because of Lemma \ref{lem:onePerProcess}, this is 
impossible.

\end{proof}
\end{theorem}
\begin{theorem}
\label{theo:accuracy}
The Algorithm presented in \autoref{fig:a1laclient1}, \autoref{fig:a1laclient2} and \autoref{fig:a1laserver}
provides \emph{accuracy}.
\begin{proof}
An accusation can be issued in lines \ref{line:a1la:convictSender}, \ref{line:a1la:proofOrigin},
\ref{line:a1la:proofAcker}.

On the first case it will have a proposal signed by a process which doesn't hold valid
signed origins for its values. One of the values can come from the process itself, in
which case a benign process would have signed it and put in its ledger on the initialisation.
The other values must come through NACKs that also provide signed origins obtained in line
\ref{line:a1la:updateLedgerProposal} which benign processes include in its ledger. Having
a proposal signed by a process which provided fake signatures to the origin of its values consist as
an irrefutable proof and $M'$ will be a non-empty subset of $M$.

The second scenario tracks processes that have inserted more than one value in the system.
A benign process would only do it once during initialisation and having two inclusions signed
by the same process consists as irrefutable proof and $M'$ will be a non-empty subset of $M$.

Finally, the decision of incomparable values requires, as seen earlier in \autoref{theo:consistency} 
that a process acknowledged incomparable values, violating the behavior of benign processes
described by Lemma \ref{lem:increasingAccept}. Having two ACKs signed by the same process
for incomparable values characterises as irrefutable proof and $M'$ will be a non-empty subset 
of $M$.

\end{proof}
\end{theorem}
\begin{theorem}
\label{theo:authenticity}
The Algorithm presented in \autoref{fig:a1laclient1}, \autoref{fig:a1laclient2} and \autoref{fig:a1laserver}
provides \emph{authentiticy}.
\begin{proof}

   This property is exactly the same as \autoref{theo:ala:authenticity}.

\end{proof}
\end{theorem}
\begin{theorem}
\label{theo:liveness}
The Algorithm presented in \autoref{fig:a1laclient1}, \autoref{fig:a1laclient2} and \autoref{fig:a1laserver}
provides \emph{liveness}.
\begin{proof}
Following \autoref{theo:validity} combined with Lemma \ref{lem:increasingProposal},
each run can have at most $\lvert U \rvert$ different values being proposed. Since by the end of this many
proposals a client shall propose the join of all these values, it will get
ACKs from a majority of processes proceeding to learn a value or gather enough information for accusing at least one byzantine
process, as at this byzantine clients must have introduced more than one value in the system for the proposal not to go through 
and \autoref{theo:accuracy} holds.
\end{proof}
\end{theorem}

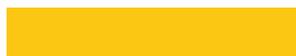
\end{document}